\documentclass[10pt]{article}
\usepackage[OE]{express}
\usepackage{cite}
\usepackage[colorlinks=true,urlcolor=blue,linkcolor=blue,citecolor=blue,breaklinks=true,bookmarks=false]{hyperref}
\usepackage{textcomp}

\usepackage{xcolor}
\usepackage[normalem]{ulem}

\begin{document}

\title{Multidimensional mode-separable frequency conversion for high-speed quantum communication}
\author{Paritosh Manurkar,$^{1,*}$ Nitin Jain,$^1$ Michael Silver,$^1$ Yu-Ping Huang,$^2$ \\Carsten Langrock,$^3$ Martin M. Fejer,$^3$ Prem Kumar,$^{1,4}$ Gregory S. Kanter$^1$}

\address{$^1$Department of Electrical Engineering and Computer Science, Northwestern University, Evanston, Illinois 60208, USA\\
$^2$Department of Physics and Engineering Physics, Stevens Institute of Technology, Hoboken, New Jersey 07030, USA\\
$^3$E.~L. Ginzton Laboratory, Stanford University, Stanford, California 94305, USA\\
$^4$Department of Physics and Astronomy, Northwestern University, Evanston, Illinois 60208, USA}
\email{$^*$paritoshmanurkar2013@u.northwestern.edu}

\newcommand{\bra}[1]{\langle #1|}
\newcommand{\ket}[1]{|#1\rangle}


\url{https://doi.org/10.1364/OPTICA.3.001300}

\begin{abstract}
Quantum frequency conversion (QFC) of photonic signals preserves quantum information while simultaneously changing the signal wavelength. A common application of QFC is to translate the wavelength of a signal compatible with the current fiber-optic infrastructure to a shorter wavelength more compatible with high quality single-photon detectors and optical memories. Recent work has investigated the use of QFC to manipulate and measure specific temporal modes (TMs) through tailoring of the pump pulses. Such a scheme holds promise for multidimensional quantum state manipulation that is both low loss and re-programmable on a fast time scale. We demonstrate the first QFC temporal mode sorting system in a four-dimensional Hilbert space, achieving a conversion efficiency and mode separability as high as $92\%$ and $0.84$, respectively. A 20-GHz pulse train is projected onto 6 different TMs, including superposition states, and mode separability with weak coherent signals is verified via photon counting. Such ultrafast high-dimensional photonic signals could enable long-distance quantum communication with high rates.\\
\end{abstract}

\textcopyright [2016] [Optical Society of America] One print or electronic copy may be made for personal use only. Systematic reproduction and distribution, duplication of any material in this paper for a fee or for commercial purposes, or modifications of the content of this paper are prohibited.

\section{Motivation}
Long distance quantum-optical communication is indispensable to several applied quantum technologies, such as quantum cryptography~\cite{bennett1984,Scarani2009}, quantum teleportation~\cite{Ma2012a}, and distributed quantum computation~\cite{Cirac1999,Ladd2010}. It also plays a vital role in the verification of fundamental tenets of physics such as tests of Bell nonlocality~\cite{Brunner2014,Shadbolt2014} and relativistic quantum information~\cite{Ralph2012}.

Ultrafast pulses~\cite{Weiner2009} produced by optical frequency combs (OFCs) with combline spacing in the $10-100\,$GHz regime routinely enable classical communication tasks such as optical networking and signal processing~\cite{Wada2004,Vlachos2009,Agrawal2010}. Their rich spectral mode structure also offers high-capacity quantum information encoding via multimode quantum states. Availability of sources and tools to exercise control over the mode structure at the transmitter, low loss and noise conditions during propagation through the optical channel, and high mode-separability and detection efficiency at the receiver then become the requisites for large-throughput quantum communication over long distances. 

One of the most extensively studied sources for high-dimensional information coding at the quantum level are the temporal modes (TMs) of ultrafast quantum states~\cite{Brecht2014,Brecht2015}. They are typically obtained by carefully engineering the process of spontaneous parametric down conversion (SPDC) or four wave mixing in nonlinear media~\cite{Eckstein2011,Eckstein2011a,Mejling2013}. If produced in the telecom wavelength regime, and in a spatial mode compatible with fibers, such multimode quantum signals are also easily integrable with the fiber-optic infrastructure. This ensures that the loss and noise incurred in propagation over long distances is low.

\begin{figure}[!t]
\centering
\includegraphics[width=0.5\linewidth]{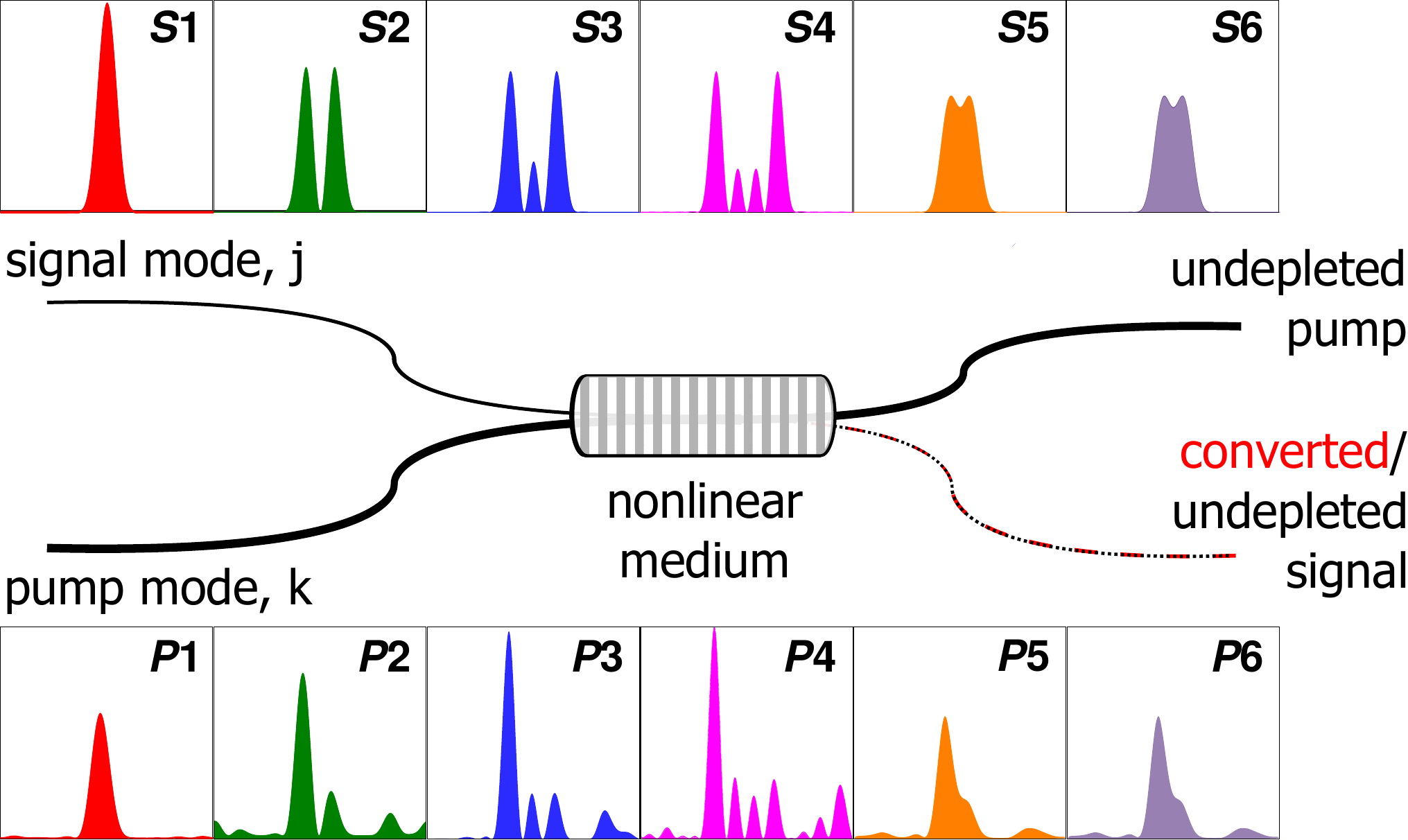}
\caption{(Color online) Mode-separable frequency conversion. The temporal modes $S1$-$S4$ were computed by simulating the SPDC process. Together with $S5$ and $S6$, superpositions of $S1$ and $S2$, these input signal pulses are converted in a mode-separable manner by appropriately shaped pump pulses $P1$-$P6$ inside the nonlinear medium. In our work, the 6x6 mode combinations are investigated as two 4x4 alphabets: $S1$-$S4$ with $P1$-$P4$, and $S3$-$S6$ with $P3$-$P6$. For an input signal $Sj$ and pump $Pk$ in a given alphabet, the signal ideally remains undepleted at the output (black-dotted line) if $j \neq k$. Otherwise, if $j=k$, its frequency gets converted (red-dashed line). This provides a way to separate the different modes. 
\label{fig:genScheme}}
\end{figure}

At the receiver, sorting of these modes is, however, not possible with only linear optics~\cite{Eckstein2011}. Furthermore, single-photon detectors (SPDs) operating in the near-infrared regime (and at non-cryogenic temperatures) typically suffer from low quantum efficiencies and high noise. Both these problems can be overcome by utilizing quantum frequency conversion (QFC), a nonlinear process in which the center frequency of the `signal' photon is changed to another value while keeping its quantum properties intact~\cite{Kumar1990}. QFC is usually realized as a sum- or difference-frequency generation process, with an intense `pump' beam providing the other frequency, as illustrated in Fig.~\ref{fig:genScheme}. 
In context of multimode quantum signals at telecom wavelengths, a \emph{mode-separable upconversion} to wavelengths in the visible regime (where highly efficient and low-noise SPDs are widely available) becomes an intrinsic property of the receiver. 

There are several QFC-based approaches for manipulating photonic temporal waveforms that permit access to high-dimensional Hilbert spaces. For instance, ultrafast-scale time delays can be mapped to measurable frequency shifts using a time-to-frequency converter~\cite{Donohue2014}. We focus on a direct manipulation of the TMs, which can be realized by tailoring the dispersion properties of the nonlinear waveguide~\cite{Eckstein2011,Reddy2013} or modulation of the pump pulses~\cite{Huang2013,Kowligy2014}. In either of these two approaches, a quantitative figure of merit based on the upconversion efficiencies $\{\eta_{kj}\}$ of the different signal modes $Sj$ with the pump being in a fixed mode $k$ is given by the separability~\cite{Reddy2013}, 
\begin{equation}
\sigma_k = \frac{\eta_{kk}}{\sum_{j=1}^{N} \eta_{kj}} ,
\label{eq:sep}
\end{equation}
where $N$ denotes the number of modes. Figure~\ref{fig:genScheme} shows the temporal profiles of signal and pump modes spanning a four-dimensional Hilbert space ($N=4$).

A unity separability ($\sigma_k = 1$) is obtained in a QFC experiment with multi-mode signals if $\eta_{kj}=0$ except for $j=k$, i.e. the pump mode $k$ converts \emph{only} signal mode $k$. In general, a more often used parameter is the selectivity $\varsigma_k=\eta_{kk}\sigma_k$, which can be understood as the probability to separate mode $k$ (in a usable manner) from all other input modes of the multi-mode signal using pump $Pk$ when measuring the generated sum frequency. However, if $\eta_{kk}$ is reasonably high, separability is a more meaningful parameter than selectivity for high-loss quantum communication scenarios (channel transmittance $T<<1$). This is because the error rate is negligible only if the separability $\sigma_{k}$ is close to unity for all $k$ while the (extra) $\eta_{kk}$ in the selectivity can simply be lumped together with all other system inefficiencies and losses, including $T$. 

Although theoretical proposals for obtaining very high conversion efficiencies and mode selectivities for quantum signals have already been made~\cite{Reddy2014,Quesada2015}, there has been limited experimental progress, especially in projecting out multiple modes. For instance, in Ref.~\cite{Brecht2014}, selectivity measurements were performed only for the fundamental pump mode. In Ref.~\cite{Kowligy2014}, a full two-dimensional Hilbert space measurement was performed but the second pump mode had a modest conversion efficiency ($\eta_{22} < 60\%$) and no superposition modes were measured. 

In this paper, we demonstrate the first multidimensional mode-separable QFC experiment, with separability measurements performed comprehensively in a four-dimensional Hilbert space. Not only do we obtain reasonably high conversion efficiencies ($\eta_{kk} > 75\%$ for any $k$) but we also show a good match between the predictions of our theoretical model and the experimental values. Also, the Hilbert space is accessed by manipulating the pump pulses using optical arbitrary waveform generation (OAWG), which notably operates only on the classical pump pulses thereby enabling fast, complex, and reprogrammable signal measurements without inserting lossy elements into the signal path. This is in contrast with the so-called orbital angular momentum states, which otherwise offer high dimensionality and manipulation with linear optics~\cite{Allen1992,Mair2001}. We also verify that the mode-separability extends to superposition states, a critical capability for quantum communications. 
\section{The building blocks}\label{sec:bblocks}
In this work, we focus on QFC realized as sum-frequency generation in nonlinear waveguides~\cite{Roussev2004,Ma2012,Brecht2013,Kowligy2014}. The conversion efficiency $\eta$ is a function of the characteristics of the waveguide and of the three participating waves: signal, pump, and sum frequency (SF)~\cite{Kumar1990}. In general, if an optical signal with average power $\rho_{\rm sig}$ yields an average SF power $\rho_{\rm sum}$, then 
\begin{equation}
\eta = \frac{\rho_{\rm sum}\lambda_{\rm sum}}{\rho_{\rm sig}\lambda_{\rm sig}} ,
\label{eq:etaWpows}
\end{equation}
with $\lambda_{\rm sum}$ [$\lambda_{\rm sig}$] denoting the SF [signal] wavelength. 

By solving the coupled-mode equations using Green's function approach~\cite{Eckstein2011,Huang2013,Reddy2013} or by calculating the unitary transformation associated with the interaction Hamiltonian~\cite{Christ2013,Quesada2015}, the input-output relations in terms of the annihilation operators of the signal and SF waves can be obtained. These relations mimic those of a multiport beamsplitter, the properties of which can be controlled by engineering the waveguide dispersion relations along with a suitable `gating' pulse --- hence the name quantum pulse gate (QPG)~\cite{Eckstein2011}. 

The introduction of QPG served as the first work that proposed and experimentally demonstrated a QFC-based mode sorting system. Note that the group velocities of the gating pulse at 870 nm and the signal in C-band in this demonstration were matched by dispersion engineering~\cite{Eckstein2011}. The gating pulse was also supposed to follow the shape of the mode to be separated. A modified implementation proposed the shaping of `arbitrary' pump pulses to manipulate the signals~\cite{Huang2013} and was based on the idea of pump pulse modulation to avoid modulation losses on the signals~\cite{Koepruelue2011}. It enabled C-band pump pulses to operate on C-band signal pulses which are compatible with the existing fiber infrastructure. The scheme was suggested for mode-resolved photon counting (MRPC) applications. Additionally, there are other schemes using multimode pumps that have also been used for manipulation of photonic signals~\cite{Pan2008,Pelc2012}.

To realize the gating or pump pulse, both QPG and MRPC approaches require synthesis of user-specified temporal waveforms. In case of ultrafast pulses, one possible solution is via OAWG, which involves \emph{shaping} the optical spectrum in the Fourier domain. For instance, the intensities and phases of the comb lines of an OFC, separated by the repetition rate of the original pulse train, may be addressed individually and independently to transform the pulse train in a desired manner. Through OAWG methods~\cite{Fontaine2011,Fontaine2012}, both the pump and signal pulses can be easily simulated as well as generated for mode-separable QFC experiments~\cite{Kowligy2014}.  
\section{Simulation and experimental tools} \label{sec:sim_exptTools}
As mentioned before, the TMs that define the `input signals' can be produced via SPDC in a waveguide. The SPDC-pump, phase matching characteristics, and the wavelength filters in the \emph{signal} and \emph{idler} after the waveguide govern the mode shapes and bandwidth of these signals~\cite{Brecht2015}. In this section, we present the main ideas behind the generation of proof-of-concept orthogonal TMs (in simulation as well as experiment). We also briefly discuss how the appropriate pump waveforms for the purpose of mode-separable QFC were then derived using these simulated signals. Both signal and pump modes were realized experimentally by shaping OFCs using OAWG. We also discuss how the scheme is able to achieve various experimental goals such as verifying the orthogonality of the input signal modes and optimizing the mode-separability in the four-dimensional Hilbert space.
\subsection{Simulation of signal and pump modes}
We numerically simulated the signal modes using techniques developed and explained in Ref.~\cite{Kowligy2014}. Briefly, we used a 30-ps-wide super-Gaussian pump pulse to drive the SPDC process and a 2.4-nm-wide square filter on the signal spectrum, producing up to 10 TMs with varying generation probabilities. In this work, we confined ourselves to the first four modes $S1$-$S4$, which also feature a relative generation probability close to one. 

Pump waveforms capable of mode-selective upconversion of signals were designed and optimized by iteratively solving the propagation equation using split-step Fourier method. This yielded the pump modes $P1$-$P4$. In agreement with the well-known limitation on maximum selectivities achievable in single-stage frequency conversion systems~\cite{Christ2013,Reddy2014}, we found that non-corresponding modes also got upconverted to some extent in our optimization. In the case of pump mode $P1$ for instance, we get $\eta_{12}=7.5\%$, $\eta_{13}=3.7\%$ and $\eta_{14}=1.5\%$. Nonetheless, with $\eta_{11}=94\%$, we obtain $\sigma_1 = 0.88$ using Eq.~\eqref{eq:sep}.

With signal mode $Sj$ corresponding to state $\ket{\psi_j}$, we computed two superposition modes $S5$ and $S6$ with $\ket{\psi_5} = 1/\sqrt{2}\left( \ket{\psi_1} + \ket{\psi_2} \right)$ and $\ket{\psi_6} = 1/\sqrt{2}\left( \ket{\psi_1} - \ket{\psi_2} \right)$, respectively. Figure~\ref{fig:genScheme} shows these modes together with their corresponding pumps $P5$ and $P6$. Although modes $S5$ and $S6$ have identical intensity profiles, their phase profiles render them orthogonal. We can classify these 6 modes into two basis sets ($S1$-$S4$ and $S3$-$S6$) since the modes in each set are mutually orthogonal.
\subsection{OFC generation, manipulation and measurement}
\begin{figure*}[!th]
\centering
\fbox{\includegraphics[width=0.8\linewidth]{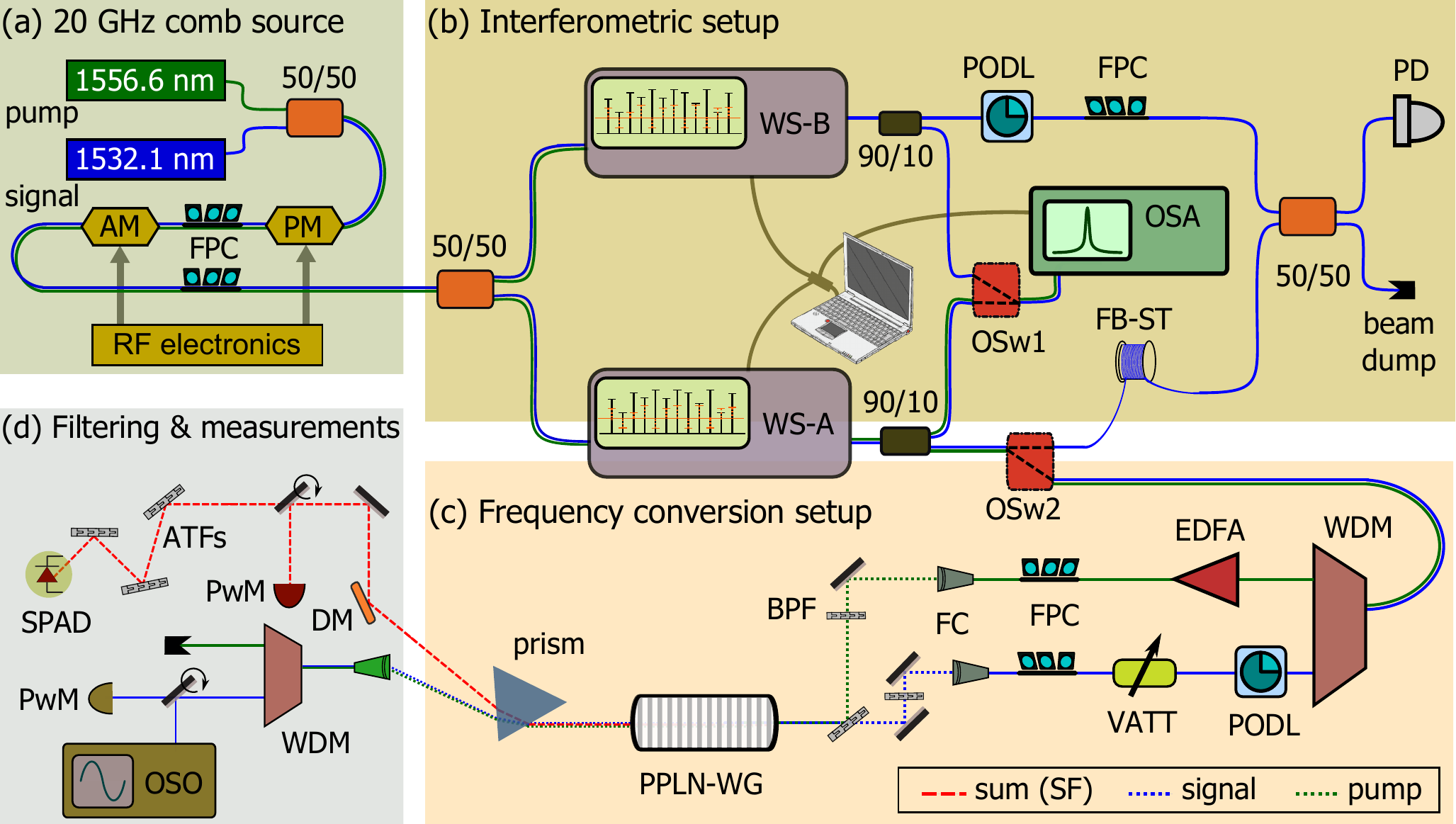}}
\caption{(Color online) Detailed optical schematic. The figure shows different interconnected setups involved in the optimal preparation of the pump and signal waveforms, various (intermediate) measurements to ensure the optimality, mode-separable frequency conversion, and the (final) power or photon counting measurements. Details of the individual components and operation of these setups are given in the main text. All solid blue/green lines denote fiber-optical paths, while dotted and dashed lines denote free-space links. FPC: fiber polarization controller, PM/AM: phase/amplitude modulator, WS: waveshaper, PODL: programmable optical delay line, OSw: optical switch, OSA: optical spectrum analyzer, FB-ST: fiber stretcher, PD: photodiode, WDM: wavelength division (de-)multiplexer, VATT: variable attenuator, WG: waveguide, PwM: classical power meter, ATF: angle tuned filter, OSO: optical sampling oscilloscope, SPAD: single photon avalanche diode.
\label{fig:setup}}
\end{figure*}

We generated two OFCs centered around the signal and pump wavelengths $\lambda_{\rm sig}=1532.1\,$nm and $\lambda_{\rm pump}=1556.6\,$nm, respectively. Each comb contained 17 lines spaced by 20 GHz. Figure~\ref{fig:setup}(a) shows the cascaded configuration of phase and amplitude modulators used for this purpose~\cite{Wu2010}. 
The actual OAWG, or line-by-line shaping to produce the signal and pump waveforms, was performed using commercial devices (Finisar WaveShaper 1000S and WaveShaper 4000S) labeled WS-A and WS-B in Fig.~\ref{fig:setup}(b). These devices essentially provide filters with programmable attenuations and phases for the entire C-band (frequency range: $191.250 - 196.275\,$THz) with a 1-GHz resolution. 

Since the combs produced using electro-optic modulators are inherently chirped, we employed various methods to remove the chirp. Previously, a fixed length of standard single-mode fiber (added after the comb source) sufficed to compensate for the chirp to the extent that some selectivity between two input modes was obtained~\cite{Kowligy2014}. However, in separate experiments, we observed some residual chirp on the pulses, which would have certainly hampered the task of mode-separable QFC in high-dimensional Hilbert space since the signal and pump waveforms $S3$-$S4$ and $P3$-$P4$ are more elaborate; see Fig.~\ref{fig:genScheme}. Thus, as described in the next section, a more exact method of chirp compensation was employed.
\subsection{Phase profile corrections}\label{sec:phaseCorrections}
For this work, we implemented a modified technique based on ideas in Ref.~\cite{Jiang2006}, in which the authors produced an arbitrary optical waveform using a comb source and a pulse shaper. For pulse characterization, they used another pulse shaper that selected only two comb lines at a time. This produced a beat signal (at the repetition rate of the pulses) which was captured on a fast photodiode and read on a sampling oscilloscope. 

Since our objective was to shape the comb phases in some desired manner rather than measure the phase profile, we adopted the same strategy but used only a single pulse-shaping device. In our case, the 20-GHz beat signal was measured on a 30-GHz sampling oscilloscope. The sinusoid measured from the first two comb lines served as the reference. As we moved to another pair, we measured the time delay of the (new) sinusoid with respect to the reference. Using the Fourier transform property that a delay in time domain is equivalent to phase-shift in the frequency domain, we calculated a suitable phase-shift that was then applied using the waveshaper to compensate for the time delay. After processing the last pair of comb lines, we obtained a chirp-free comb (flat phase profile) by applying the appropriate `phase corrections' to each comb line. 

In case of signals $S1$-$S6$, the simulated phase values consist of either $0$ or $\pi$, which were programmed directly in the experiment. The pump waveforms, however, had more complex spectral phases than signals. The program therefore simply produced flat phase profile for pump, and the target phase profile was applied later --- as the sum of the simulated phase values and the phase corrections --- via the pulse shaper. Note that this method of correcting phases assumes that there would be negligible impact of amplitude changes on the measured phase values.
\subsection{Interferometric rectification of signals' orthogonality}\label{sec:interferometer}
In theory, the simulated SPDC modes in the two sets $S1$-$S4$ and $S3$-$S6$ are perfectly orthogonal to each other. However, due to experimental imperfections and errors, the overlap integral of two dissimilar signals drawn from the set may not always be zero. With features on a picosecond scale, even subpicosecond time-shifts between the centers of the two signals can lead to non-negligible mode overlap. This would have an adverse impact on the mode separabilities in the QFC experiment. 

In order to rectify this, we designed a Mach-Zehnder interferometer containing pulse shapers WS-A and WS-B in its two arms, as shown in Fig.~\ref{fig:setup}(b). With one of the signals chosen as a reference on WS-B, we shaped the four signals on WS-A one by one. The delay between the two shaped signals was set using a programmable optical delay line (PODL) while the interference fringes were produced by driving a homemade fiber stretcher (FB-ST) with a triangular wave. One could calculate the interference visibility at the given delay from these fringes.

Figure~\ref{fig:visibNpumpopt}(a) shows such experimental visibility-versus-delay patterns for $S4$, $S3$, and $S6$ on WS-A and reference $S3$ on WS-B. The origin or the `0' delay is determined by the $S4$-$S3$ measurement, and it can be observed that the experimental points match well with the theoretical curve (black dashed line). Before rectification, the $S3$-$S3$ maxima and the $S6$-$S3$ minima (orange and purple traces) are temporally shifted by 1.3 ps and 1.8 ps, respectively, with respect to the origin. Thus, we shifted $S3$ and $S6$ by applying appropriate phase-shifts via WS-A to improve orthogonality. We typically obtained a visibility $<0.04$ at the origin for any two dissimilar signals after such an alignment. 
\subsection{Pump phase optimization}\label{sec:pumpOpt}
High-quality mode separability, in practice, relies on precise shaping of the numerically simulated pump waveform. The pump in the experiment may, however, differ from ideal due to imperfections in the pulse-shaping process. Furthermore, non-ideal phase matching in the waveguide (which is not accounted for in the simulation) can worsen the mode-separability. 
\begin{figure}[!b]
\centering
\includegraphics[width=0.5\linewidth]{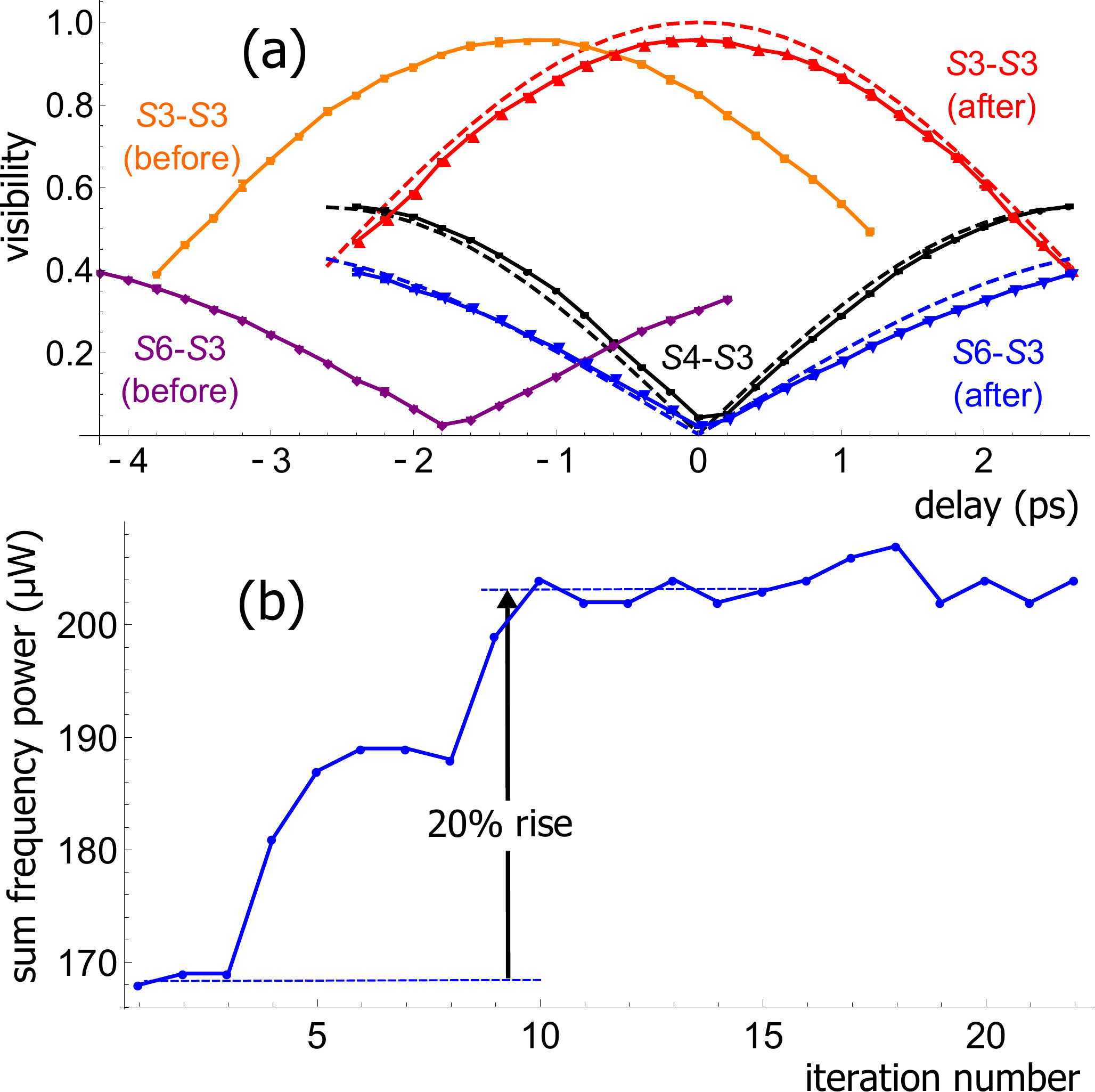}
\caption{(Color online) Interference visibility vs delay and pump-phase optimization. (a) Interference patterns with $S3$ as the reference signal on WS-B. The delay value on the PODL that yields the lowest $S4$-$S3$ visibility sets the origin. After the interferometric rectification, the measurements show good match between the theoretical (dashed) and experimental (solid) curves. (b) By means of optimizing the phase of the 17 comb lines of pump $P3$, we obtain around $20\%$ increase in $S3$ conversion, which is directly related to $\eta_{33}$ per Eq.~\eqref{eq:etaWpows}.
\label{fig:visibNpumpopt}}
\end{figure}
To mitigate such effects, we developed an experimental pump optimization technique based on the simultaneous perturbation stochastic approximation (SPSA) algorithm~\cite{Spall1992}. 

Given an imperfectly prepared pump $Pk$, the aim of the optimization was to increase the conversion efficiency $\eta_{kj}$ of signal $Sj$ if $j=k$, and decrease it otherwise. This involved perturbing the pump and monitoring the changes in the SF power (directly proportional to $\eta_{kj}$, see Eq.~\eqref{eq:etaWpows}) with the power meter shown in Fig.~\ref{fig:setup}(d). The optimization was performed in a feedback loop involving WS-A (on which the pump comb $Pk$ was shaped iteratively along with the signal $Sj$), the SF power meter, and a computer communicating with both these devices. 

The SPSA algorithm is suitable for noisy multi-variable systems and employs a gradient approximation to steer the next choices of the variables. Each pump mode $P1$-$P6$ is defined by 34 variables --- 17 amplitudes and 17 phases --- that could as such be perturbed simultaneously for the purpose of gradient estimation. However, manipulations in such a large space would render the optimization quite slow. Since we could measure the comb line intensities on an optical spectrum analyzer (OSA) and ensure that they stay within an error margin from the target, we decided to limit the pump optimization process to the 17 phase values only. Further details of the algorithm and the implementation will be discussed elsewhere (unpublished data). 

It is not possible to predict the number of iterations required for convergence due to the stochastic nature of the SPSA algorithm. We let the optimization run until the SF power appeared to reach a consistent maximum/minimum. Typically, this happened within 10-40 iterations. Figure~\ref{fig:visibNpumpopt}(b) shows an example where the conversion of $S3$ from $P3$ was optimized. One can observe that the SF power (blue trace) evolves rapidly until the tenth iteration, after which it fluctuates around a mean value just above $200\,\mu$W. Note that the optimization can also be used to lower a conversion efficiency $\eta_{kj}$ when $j \neq k$. In general, after performing any such optimization with pump $Pk$, we rechecked the separability $\sigma_k$ to ensure it also improved. 
\section{Mode-separable frequency conversion}\label{sec:exptDetails}
Figure~\ref{fig:setup} shows the experimental schematic, parts (a) and (b) of which have been discussed already in Sec.~\ref{sec:sim_exptTools}. Briefly, part (a) shows the setup for generating OFCs centered at $\lambda_{\rm sig}$ and $\lambda_{\rm pump}$ while the interferometric setup in part (b) was employed to ensure any two distinct signals in the set $S1$-$S4$ (or $S3$-$S6$) were as orthogonal as possible. Since the interferometric measurements were performed only on the signals, both WS-A and WS-B were controlled to suppress transmission around $\lambda_{\rm pump}$. The reference signal $Sk$ was shaped via WS-B while signal $Sj$, chosen from the appropriate set, was shaped on WS-A. 
\subsection{Details of the setup}\label{sec:detailsFCsetup}
Once the signals' orthogonality were rectified interferometrically, we sent the signal and pump to part (c) of Fig.~\ref{fig:setup} by changing the state of an optical switch (OSw2) and shaping pump $Pk$ (along with signal $Sj$) on WS-A. There, we first separated the co-propagating pump and signal, amplified the pump using a 3-W Er-doped fiber amplifier (IPG Photonics), and tuned the relative delay between pump and signal in order to make sure they overlapped temporally inside the waveguide. Using a variable attenuator (VATT), the intensity of the signal $Sj$ could be attenuated, if required, to well below the single-photon regime, i.e., mean photon number $\mu_j << 1$ per pulse. The shaped and optimized pump $Pk$ and signal $Sj$ were filtered (to suppress ASE) and combined in free space using an angle-tuned filter (ATF) before the waveguide.

The frequency conversion was performed in a 52-mm long periodically poled lithium niobate (PPLN) waveguide, fabricated at Stanford University~\cite{Parameswaran2002}. The waveguide featured a clean single phase-matching peak centered at approximately $1544\,$nm at $73.4 \,^{\circ}\mathrm{C}$, and a small-signal second-harmonic generation efficiency of $1600\%$/W. The pump wavelength of $\lambda_{\rm pump}=1556.6\,$nm was chosen to be far away from the phase-matching peak in order to suppress any unwanted second harmonic (SH) from the pump. Through subsequent measurements of the SF phase-matching curve using CW signal and 160-ps pump at 50 MHz, we found the highest SF conversion efficiency of $93.6\%$ at a signal wavelength $\lambda_{\rm sig}=1532.1\,$nm and peak pump power $\approx 94\,$mW.  

We separated the pump and (depleted) signal beams from the SF beam (center wavelength $\lambda_{\rm sum}=772.1\,$nm) at the output of the waveguide using a prism. As shown in Fig.~\ref{fig:setup}(d), we could direct the SF towards a power meter for classical measurements or towards a single-photon counting module (SPCM model, AQRH-14) setup with a flip mirror. The SPCM setup employed a series of ATFs to get rid of photons generated at the SH wavelength of $\lambda_{\rm pump}/2$. 

We also separated the signal from pump using a coarse wavelength division demultiplexer centered at 1531 nm and observed its depletion on a power meter. The signal waveform could also be observed on a 500-GHz optical sampling oscilloscope. 
\subsection{Mode-separable measurements}\label{sec:procedure}
Let us summarize the experimental procedure for performing mode-separable measurements assuming the pump to be fixed in mode $P1$. We first corrected the phase profiles for each of the signals in the set $S1$-$S4$ as described in Sec.~\ref{sec:phaseCorrections}. The phase correction procedure for $P1$ produced a flat profile, to which the simulation-determined phase values were then added. We then shaped $S1$ as reference on WS-B (kept untouched from hereon) and performed interferometric measurements while shaping $S1$-$S4$ on WS-A one at a time, as described in Sec.~\ref{sec:interferometer}. These measurements allowed us to align all four signals in time by correcting for any relative shifts determined with respect to a common reference, as explained via Fig.~\ref{fig:visibNpumpopt}(a).  

With $P1$ and $S1$ shaped on WS-A, we then varied the relative delay $\delta_{11}$ between the signal and pump pulse trains, and the average pump power $\rho_1$ entering the waveguide to maximize $\eta_{11}$, calculated using Eq.~\eqref{eq:etaWpows}. This gave us an optimal delay $\delta^{\rm opt}_{11}$ and optimal average pump power $\rho_1^{\rm opt}$ at which we measured the conversion efficiencies $\eta_{1j}$ of signals $Sj$ for $j = [2,4]$. From this, we could calculate the separability $\sigma_1$ using Eq.~\eqref{eq:sep}. 

Next we optimized $P1$ phase, as described in Sec.~\ref{sec:pumpOpt}, to reduce the worst-performing conversion efficiency $\eta_{1j}$ (in relation to the simulated value). With this stochastically optimized pump phase, we repeated the measurements of $\eta_{1j}$ for $j = [1,4]$ at the same pump power $\rho_1^{\rm opt}$ and re-calculated $\sigma_1$. This process was repeated a few times to find the final pump phase profile, i.e., whichever yielded the highest $\sigma_1$. 
\begin{figure}[!ht]
\centering
\includegraphics[width=0.65\linewidth]{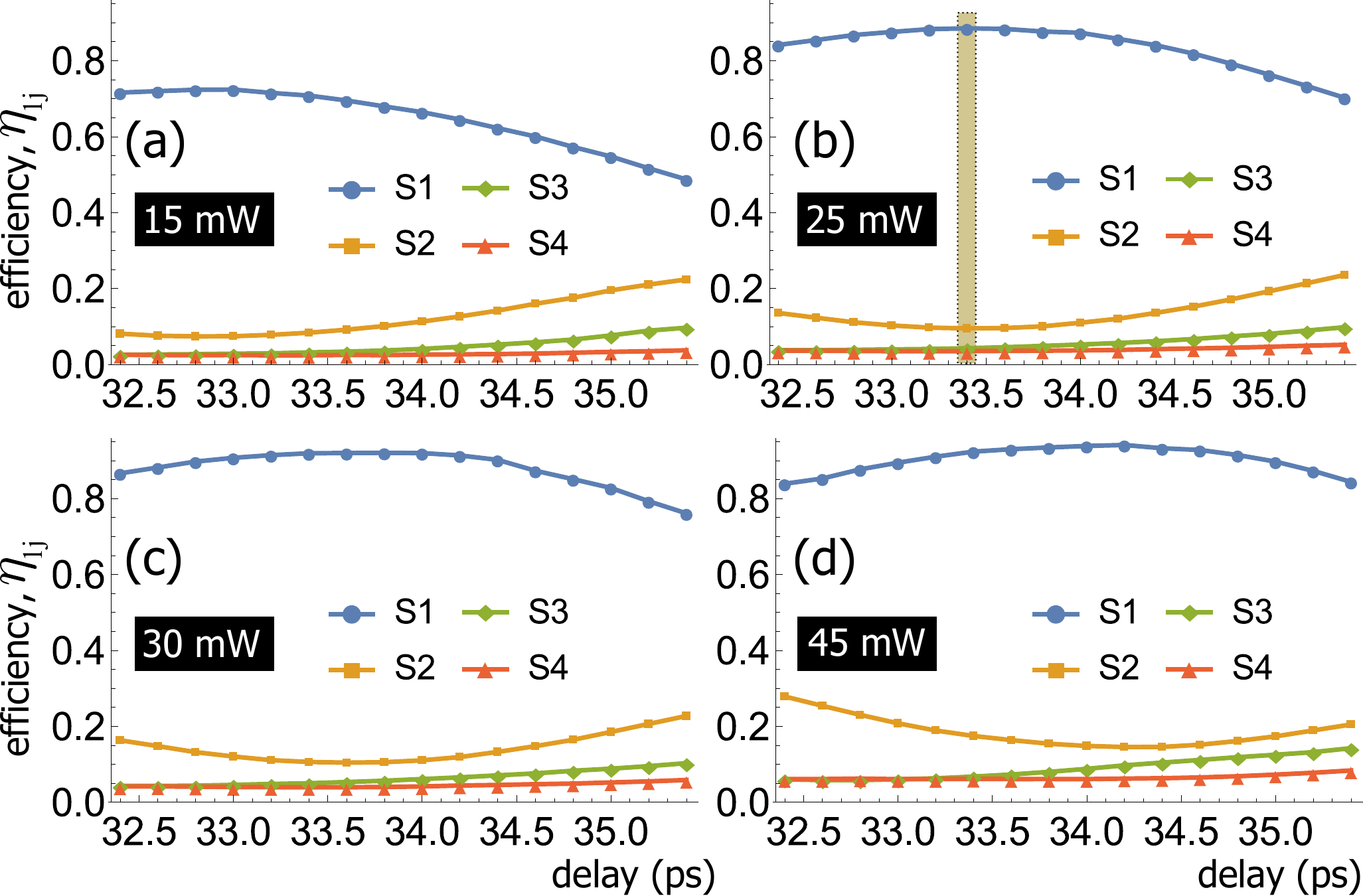}
\caption{(Color online) Conversion efficiencies of signals $S1$-$S4$ measured at various average pump powers and pump-signal delay. For all pump powers (indicated in boxes), the delay was varied in steps of $0.2\,$ps. The maximum separability was obtained at an average pump power $\rho^{\rm fin}_1 = 25\,$mW and delay $\delta^{\rm fin}_{11} = 33.4\,$ps, indicated by the shaded rectangle in (b). Error bars are included but may be too small to be observed. 
\label{fig:etaP1vsDelnPow}}
\end{figure}
Figure~\ref{fig:etaP1vsDelnPow}(b) shows an example of the measurement results from which $\sigma_1=0.84$ was calculated. 

For measurements with the SPCM setup, we attenuated the signal arm to mean photon numbers $\mu_j \approx 0.15 \pm 0.05$ per pulse exiting the waveguide. Due to the 20-GHz system repetition rate, we introduced an additional loss of $\sim18\,$dB in the SF path to prevent saturating the SPCM and limit the photon-count rates to $<10\,$Mcps. Finally, we repeated the interferometric measurements with the unchanged $S1$ on WS-B and $S1$-$S4$ re-shaped on WS-A to verify the stability of the signals' orthogonality.
\subsection{Results}\label{sec:results}
The ultimate goal of the experiment was to shape the two 4x4 alphabets shown in Fig.~\ref{fig:genScheme} in an optimal manner and measure the conversion efficiencies $\eta_{kj}$ for all relevant combinations of pump $Pk$ and signal $Sj$ ($j,k = [1,6]$). For each $Pk$-$Sj$ combination, we performed multiple power and delay measurements in steps of 5 mW (average power) and 0.2 ps around the previously determined optimal values $\rho_k^{\rm opt}$ and $\delta^{\rm opt}_{kk}$, respectively. As illustrated in Fig.~\ref{fig:etaP1vsDelnPow} for pump $P1$, we calculated $\eta_{kj}$ at each delay and power. Subsequently, the separability $\sigma_k$ was evaluated using \eqref{eq:sep}. The delay $\delta^{\rm fin}_{kk}$ and power $\rho_k^{\rm fin}$ at which $\sigma_k$ was maximum were used for plotting the final data. The optimal and final values were typically the same. 

\begin{figure}[!t]
\centering
\includegraphics[width=0.6\linewidth]{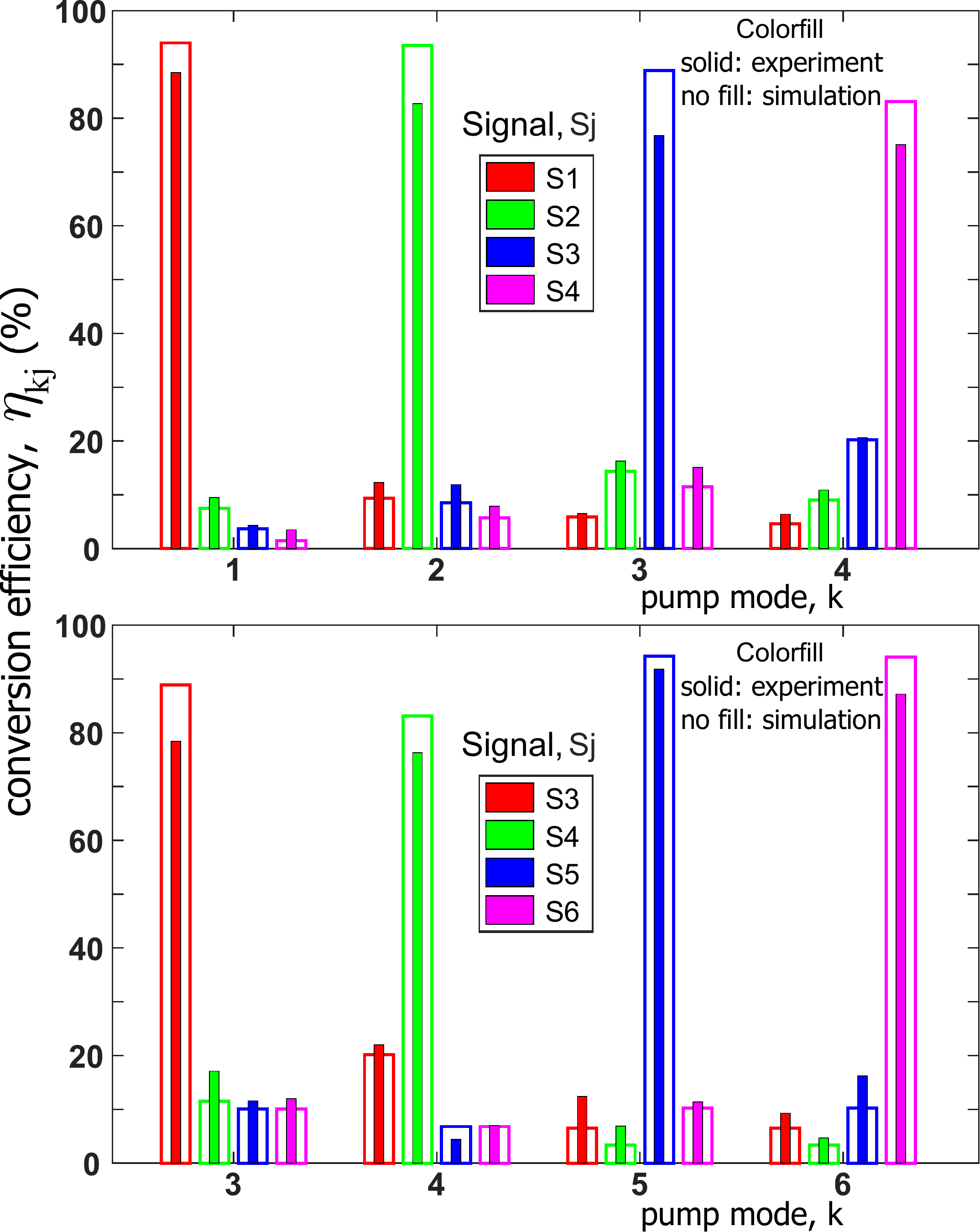}
\caption{(Color online) Conversion efficiency results from the two 4x4 alphabets. For any given pump $Pk$, the average pump power and relative delay between the pump and signal pulses in the experiment were the same across all four signals $Sj$. 
\label{fig:both4x4}}
\end{figure}

Figure~\ref{fig:both4x4} plots the simulation-derived $\eta_{kj}$ values along with experimental results with classical signals at delay $\delta^{\rm fin}_{kk}$ and power $\rho_k^{\rm fin}$. 
The mode-selective upconversion of $S5$ and $S6$ by their respective pumps $P5$ and $P6$ clearly shows our experimental capability to measure superposition states, as is commonly required in quantum communications. 

\begin{figure}[!t]
\centering
\includegraphics[width=0.5\linewidth]{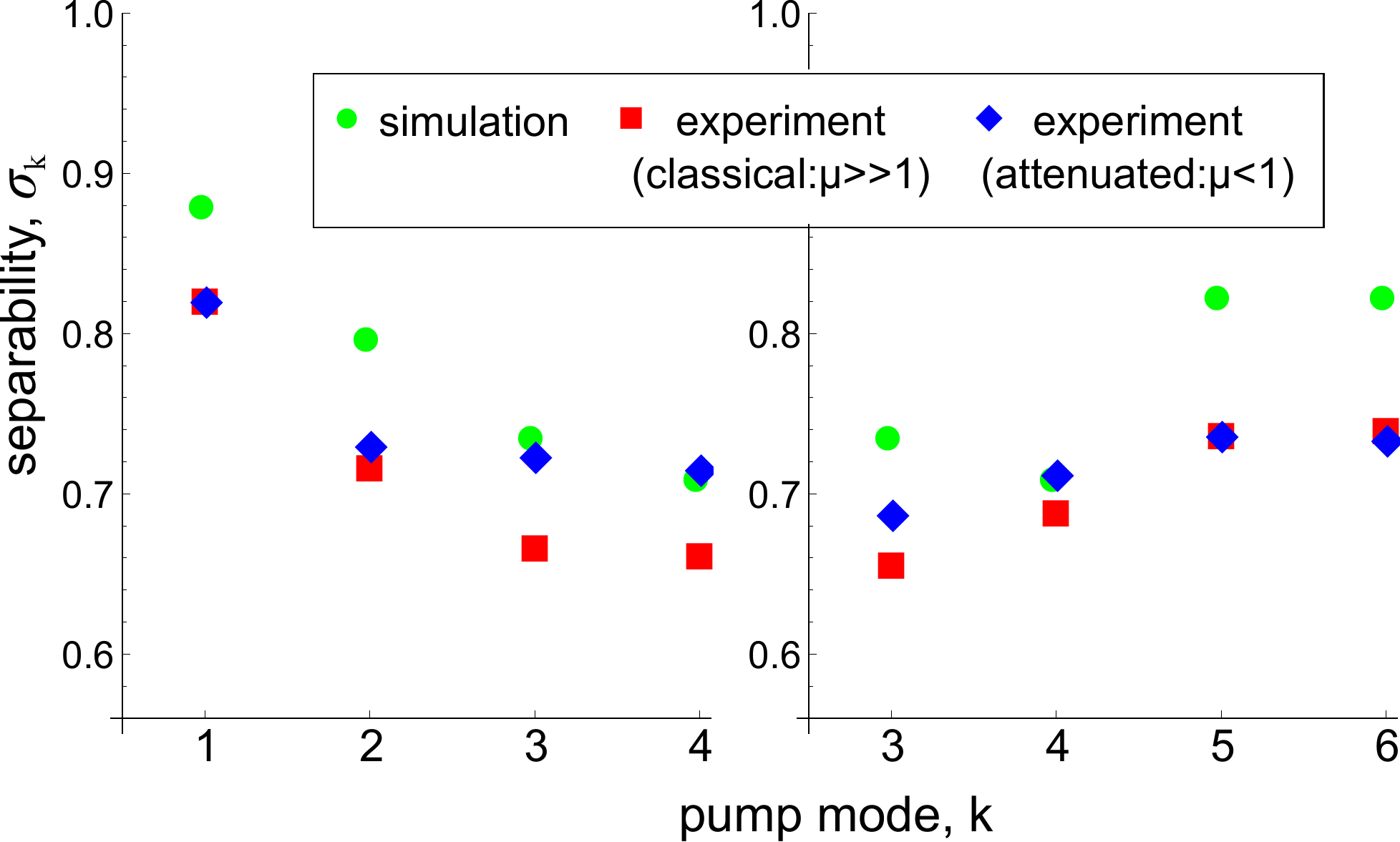}
\caption{(Color online) Experimental separability results with classical ($\mu \gg 1$) and weak coherent ($\mu < 1$) signals $S1$-$S6$ that were upconverted by pump waveforms $P1$-$P6$ in a mode-separable manner. These results were obtained at a pump-signal delay $\delta^{\rm fin}_{kk}$ and pump power $\rho_k^{\rm fin}$. Separabilities evaluated using the simulation model are also shown.
\label{fig:sepall}}
\end{figure}

Figure~\ref{fig:sepall} shows the final separabilities $\sigma_k$ with $k = [1,4]$ for the first 4x4 alphabet and $k = [3,6]$ for the second 4x4 alphabet in both experiment and simulation. For the classical case, the calculations were done using the conversion efficiencies plotted in Fig.~\ref{fig:both4x4}. We also ascertained $\sigma_k$ using photon counting, as outlined in the previous section. To account for noise due to pump-induced processes such as parametric fluorescence and spontaneous Raman scattering~\cite{Pelc2011}, we also separately recorded the counts by physically blocking the signal before the waveguide; see Fig.~\ref{fig:setup}. The average SF photon generation rate with both $Pk$-$Sk$ propagating through the waveguide to that of noise photons (i.e., with only $Pk$ propagating) was $>10^3$, which is more than two orders of magnitude higher than the signal-to-noise ratio (SNR) reported in Ref.~\cite{Brecht2014} at the same $\mu_j \approx 0.15$. 

The noise-corrected separability results are shown in Fig.~\ref{fig:sepall}. They are reasonably consistent with the classical/simulation results, thereby demonstrating mode-separable QFC in the four-dimensional Hilbert space. Note that using lower pump powers can dramatically reduce the noise counts while affecting $\sigma_k$ rather insignificantly. For instance, decreasing the average pump power of $P6$ from 35 mW to 25 mW reduces signal counts by $<8\%$ while the decrease in noise counts is $>40\%$. Alternatively, longer wavelength pumps ($\lambda_{\rm pump} > 1800\,$nm~\cite{Pelc2011}) also result in less noise, though commercial OAWG technology is not currently available at such wavelengths. 
\section{Discussion}
In principle, TMs span an infinite-dimensional Hilbert space, which, in practice, implies higher channel capacity per photon if the quantum information can be conveniently encoded and decoded. If the coding encompasses modal superpositions, these modes are highly relevant for tasks such as quantum key distribution (QKD) and quantum state tomography~\cite{Brecht2015,Velev2014}. High-dimensional QKD protocols, apart from offering enhanced key rates, may also increase the security against eavesdropping~\cite{Cerf2002}.

The system design shown in Fig.~\ref{fig:setup} is essentially that of a communication receiver with the classical/quantum signals' source connected to it with a very short channel. From the perspective of quantum communication, one can instead imagine the quantum signal source and interferometric setup to be connected to the receiver by fiber-optic channels (along with a shared clock). Assuming the loss parameter as $0.2\,$dB/km, high-dimensional mode-separable photon counting with off-the-shelf photon counting systems~\cite{MPDspcm} would be possible with our receiver for links in the $100\,$-km range. This is because even after the $20\,$dB loss penalty, the SNR (initial value $>10^3$) is at least $10$, and could also be improved by methods discussed above. For shorter links, one can take advantage of the 20-GHz repetition rate to achieve high throughputs by employing ultralow-jitter single-photon detectors~\cite{Thew2006,Ma2015}. This indicates the feasibility of QKD which we discuss below via a numerical example based on the analysis in Ref.~\cite{Brecht2015}. 

We assume Alice's source capable of generating single photon or weak coherent pulses with $\mu < 1$ (as used in Sec.~\ref{sec:results}, and as is also the case for many QKD protocols) in state $\ket{\psi_j}$ of the signal mode $Sj$.
The subset consisting of $\ket{\psi_1}$, $\ket{\psi_2}$, $\ket{\psi_5} = 1/\sqrt{2}\left( \ket{\psi_1} + \ket{\psi_2} \right)$ and $\ket{\psi_6} = 1/\sqrt{2}\left( \ket{\psi_1} - \ket{\psi_2} \right)$ forms a BB84 alphabet~\cite{bennett1984}. These states can be measured in a mode-separable manner by employing pump modes $P1$ and $P5$; for instance, if Alice prepares and sends $\ket{\psi_5}$ and Bob employs $P5$, he obtains the correct measurement outcome with a probability proportional to $\eta_{55}$. If however, Alice's state had been $\ket{\psi_6}$, Bob would infer the wrong bit with a probability proportional to $\eta_{56}$. 

Using the data shown in Fig.~\ref{fig:both4x4}, we find the separabilities $\sigma^{(2)}_1 = 0.903$ and $\sigma^{(2)}_5 = 0.890$; the superscript $(2)$ here denotes that the calculation is over a 2x2 alphabet. Assuming an asymmetric BB84 with $P1$ for monitoring the quantum bit error ratio (QBER) and $P5$ for the raw key generation, the overall receiver efficiency is $\eta_{\rm ov} \approx \eta_{55}\sigma^{(2)}_5 = 81.7\%$ (assuming negligible waveguide/coupling loss and unity detection efficiency). The high system repetition rate would allow significant raw key rates even in case of very lossy channels. The contribution to the QBER $\approx 1-\sigma^{(2)}_1 = 9.7\%$ is also high. But this can be brought down with improved separabilities, possible through experimental advances including larger OAWG bandwidths and longer waveguides, e.g., using 30 comb lines separated by 40 GHz in a 6-cm-long waveguide~\cite{Kowligy2014}. The QBER could be made negligible with even more elaborate receiver configurations that promise unity separabilities~\cite{Reddy2014}.  

Alternatively, with modes $S1$-$S4$ and $P1$-$P4$, one can operate a genuine four-dimensional QKD protocol (basis size $d=4$). The tasks of random selection from $d(d+1)=20$ possible basis states at the source and $d-1=3$ measurements at the receiver (to distinguish in each of the $d$ bases~\cite{Brecht2015}) could be efficiently carried out with dynamic OAWG capabilities and low-jitter detectors~\cite{Fontaine2011,Fontaine2012,MPDspcm,Thew2006}. Although the contribution to the QBER from non-unity separabilities will be higher than in the BB84 example above, high-dimensional QKD protocols also generally have higher QBER tolerances~\cite{Cerf2002}. 

Finally, let us note the pros and cons of using TMs in comparison to orbital angular momentum (OAM) states~\cite{Allen1992,Mair2001}, which offer encoding of quantum information in very high-dimensional Hilbert spaces with linear optics. The primary advantage of TMs is that with dynamic OAWG capabilities~\cite{Fontaine2011,Fontaine2012}, even fairly complex manipulations of the quantum signal can be done at very low loss because the mode manipulation (OAWG) is performed on the pump beam instead of the signal beam. Additionally, OAWGs have been demonstrated with extremely large update bandwidths. Also, TMs can be easily produced at telecom wavelengths and in spatial modes that make them highly compatible with the existing fiber-optic infrastructure. 
\section{Conclusion}
In conclusion, we have demonstrated a frequency conversion based system for sorting temporal modes chosen from a four-dimensional Hilbert space. Using optical arbitrary waveform generation, we have shaped pump and signal pulses that undergo sum frequency generation in a nonlinear waveguide. By employing novel experimental techniques to verify/optimize the mode quality of the signal/pump, we have achieved high conversion efficiencies and separabilities for all modes, including superposition states. Apart from enabling high-dimensional, long-distance quantum communication, mode-separable frequency conversion also provides elegant and efficient solutions to problems such as quantum-state purification and mode reshaping of single photons, which have applications in quantum networks and hybrid interfaces.
\section*{Funding Information}
Defense Advanced Research Projects Agency (DARPA) Quiness program (W31P4Q-13-1-0004).
\section*{Acknowledgment}
This research was supported in part by the DARPA Quiness program (Grant \# W31P4Q-13-1-0004). It should be noted that although Prem Kumar managed the Quiness program in his capacity as a DARPA Program Manager, the Northwestern effort was delegated to another Program Manager to avoid conflict-of-interest issues.

\bibliographystyle{plain}
\bibliography{library,misc}

\end{document}